\documentclass[debug,overfull]{epl} 
\usepackage{graphicx}
\usepackage{dcolumn}
\usepackage{amsmath}
\usepackage[latin1]{inputenc}
\title{The energy landscape of aging systems---from a different angle}
\shorttitle{Energy landscape of aging systems}
\author{Paolo Sibani \and Jesper Dall} 
\institute{Fysisk Institut, Syddansk Universitet, 
DK--5230 Odense M, Denmark} 
\begin{document}
\date{\today} 
\maketitle
\begin{abstract}
A novel method for  glassy landscape exploration 
is presented which utilizes 
a time series of energy values 
collected during an isothermal relaxation  after  a 
thermal quench.
A sub-series of increasingly rare events, or quakes, 
which are connected to an irreversible release 
of energy from the system, 
is used to identify  entry and exit times 
for landscape valleys. The landscape 
of three dimensional spin glasses is studied from this angle for 
a number of lattice sizes and for a range of low temperatures.
A simple   picture emerges regarding the temperature and size
dependence of 
(1) the energy barriers separating the valleys, 
(2) the lowest energy minimum within a valley, and 
(3) the distance between the configurations belonging to 
the lowest minima in neighboring valleys. 
The     configuration  changes following the quakes
are analyzed in terms of connected clusters of flipped  spins, 
 and the  size distribution of these clusters is presented.

 \end{abstract}

\section{Introduction} 
Macroscopic properties 
of thermalizing glassy systems depend on the time (age) 
elapsed after the quench into the glass
phase: on time scales shorter than 
 the age, pseudo thermal equilibrium 
 establishes itself  locally within metastable regions of the landscape 
 while on longer observational time scales
the relaxation is manifestly 
non-stationary~\cite{Alba86,Svedlindh87,Andersson92}.
 Memory and rejuvenation effects
 resembling those observed in 
 spin glasses~\cite{Jonason98}
 are also present in e.g.\ microscopic models 
 of driven dissipative systems~\cite{Sibani01}, where 
they are related to an irreversible dynamical selection of marginally stable 
attractors~\cite{Coppersmith87,Sibani93a,Sibani01}. This 
 suggests that similar mechanisms could be present 
in thermal aging as well.
 
To discuss the connection between aging and landscape
geometry it is crucial 
(\emph{i}) to identify the dynamical events, or 'quakes', marking
 the transition between 
equilibrium and non-equilibrium dynamics, and 
(\emph{ii}) to characterize in 
 configuration and/or real space the attractors,
 or 'valleys' 
 \emph{dynamically selected} by these quakes.
 The approach developed below 
strives to guarantee the dynamical relevance 
of the structures identified (minima, 
barriers, etc.) by solely 
relying on a statistical analysis of 
time series collected during unperturbed 
aging following an initial quench.
Our method  is generally applicable to glassy 
systems and can, with a small computational overhead, complement 
established landscape mapping procedures such as e.g.\ the Stillinger-Weber
approach~\cite{Stillinger83,Nemoto88,Becker97,Crisanti02,Mossa02,Doliwa03},
which employ e.g.\ thermal quenches to study local energy minima.

The method is  demonstrated by 
an application to short range spin glasses, mainly
nearest neighbor models on three dimensional cubic lattices. 
Our results concern 
the energy barriers which are surmounted by the quakes, 
 the frozen-in  energy they   release, 
and  the corresponding configurational 
rearrangements.

 \begin{figure}[ht!]
\twofigures[scale=0.37]{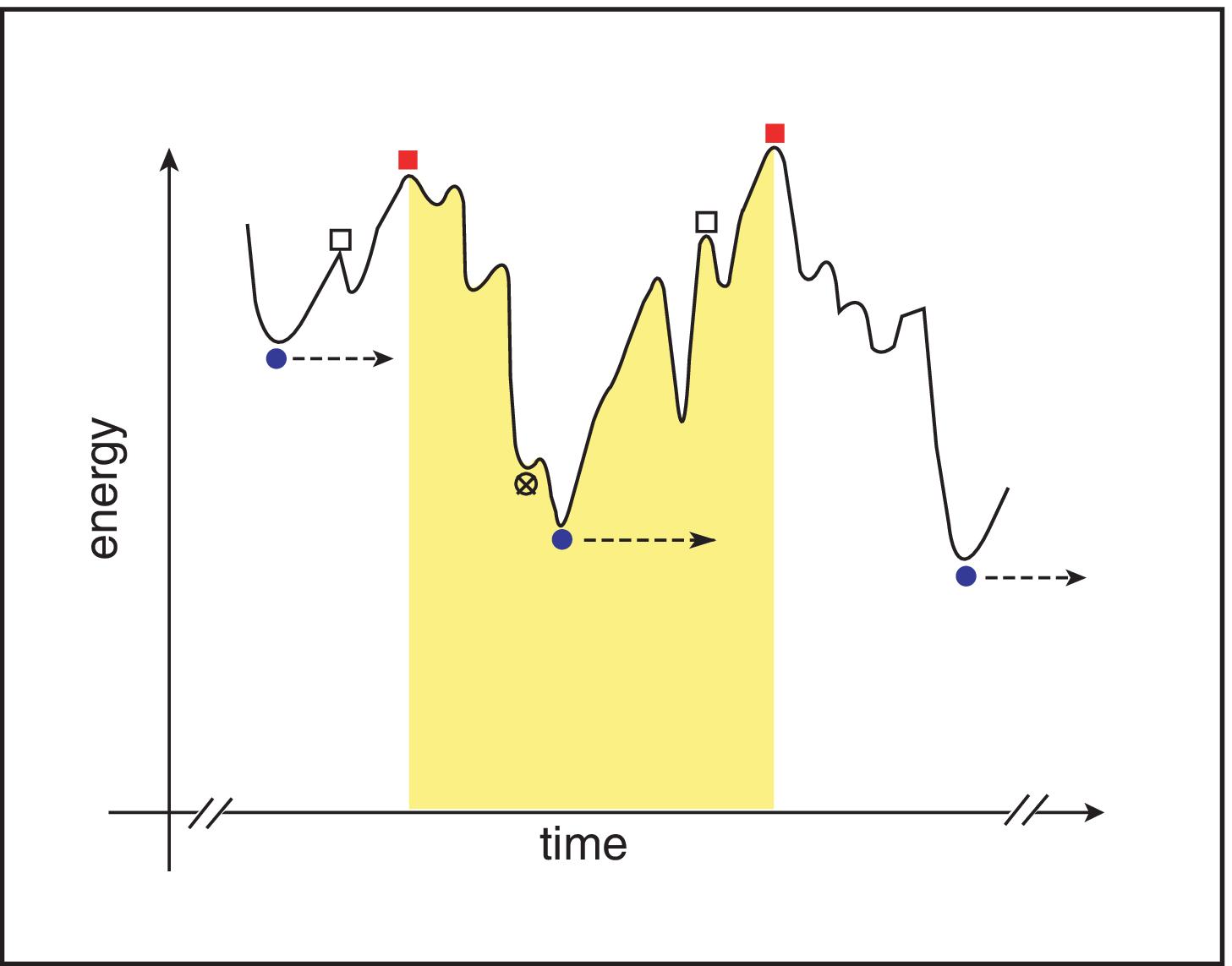}{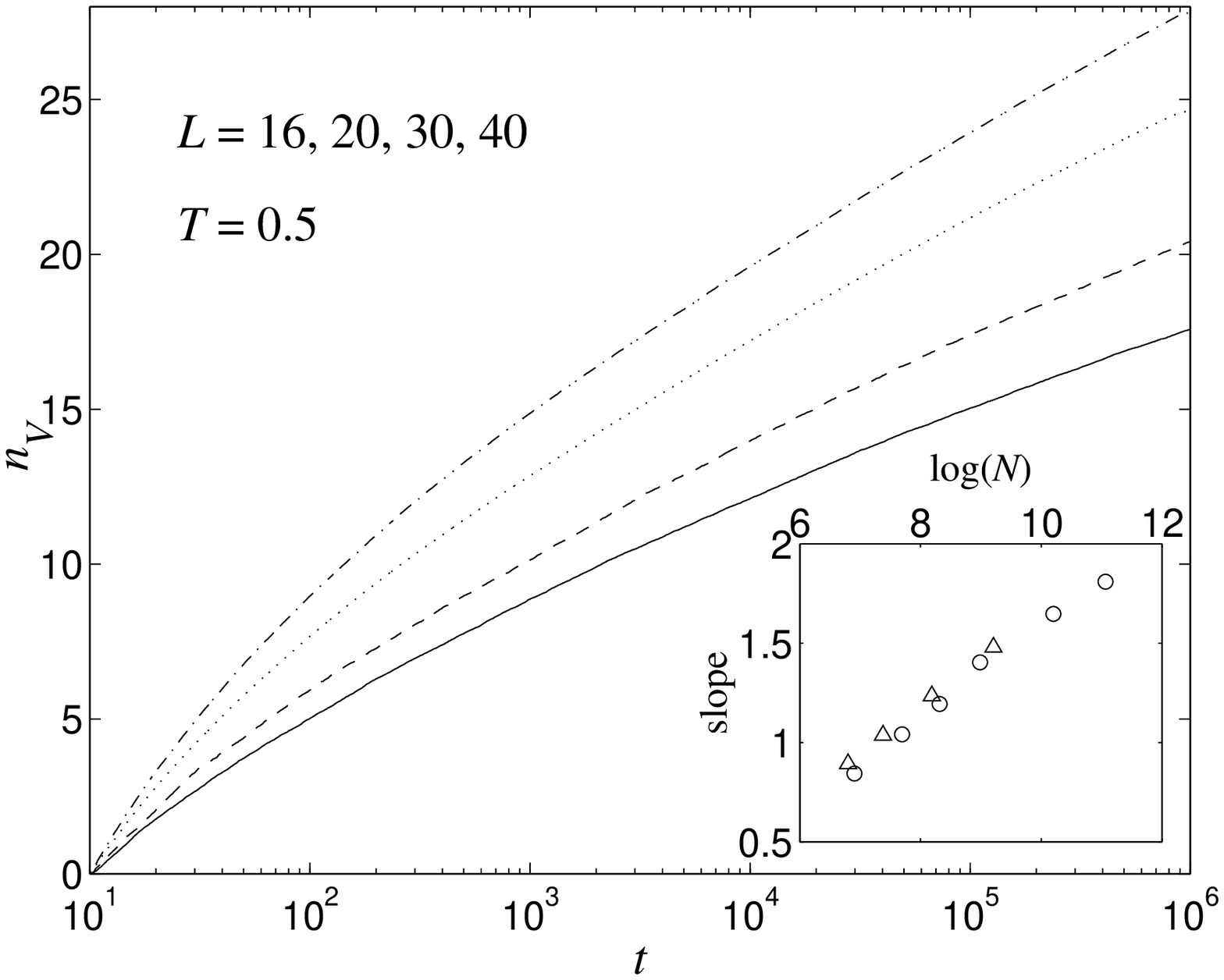} 
\caption
{\small A fictitious time series of collected energy values 
illustrates our sampling procedure and our definition
of a valley (= gray area). Conspicuous events, i.e.\ 
energy and barrier records, are denoted by circles and squares.
The black squares indicate the entry and exit time
in the valley, while the black circles show the lowest energy record
which allegedly corresponds  to an  energy minimum. 
}
\label{fig1} 
\caption{ \small Main figure: the average number $\overline{n_V}$ 
of valleys visited in the time interval $[0,t]$
is plotted versus time. The data pertain to $3d$ gaussian spin glasses of size $N=L^3$.
Insert: the corresponding logarithmic slopes of $\overline{n_V}$ as a function
of the logarithm of the system size $N$. These data are nearly insensitive to 
temperature or system type: 
Triangles: $2d$, $T=0.4$. Circles: $3d$, $T=0.5$.
}
\label{B_in_time} 
\end{figure}

\section{Method} 
The idea that progressively deeper and more stable attractors are explored 
during aging appears in different guises in a number of 
mesoscopic   models~\cite{Sibani89,Lederman91,Sibani91,Joh96,Hoffmann97,Sibani97a,Joh99} 
which can reproduce many aspects of aging phenomenology.
Implicit in that sort of modeling is a
temporal sequence of increasingly rare non-equilibrium 
events each 
overcoming a dynamical barrier higher than all
those previously surmounted.
 
To investigate the occurrence of such events 
in the actual dynamics of microscopic models, 
it is natural to look at extremal values 
of the energy in the data stream 
produced by a simulation: 
We keep track of the current lowest energy 
encountered, $E$, and of the current highest barrier $B$
overcome. With a slight abuse of notation,
a barrier is simply  the energy of 
the current state measured from $E$, and, as such,  does not 
 necessarily  provide access to new
regions of configuration space. Henceforth,
$E$ and $B$ will be referred to as 
energy and barrier records. 

Whenever a system is 
locally equilibrated at low $T$, the current state of lowest 
energy is likely to be repeatedly visited by the 
(recurrent) fluctuations. By this token, 
the appearance of a lower $E$ value 
 implies that local equilibrium 
is broken. Once a trajectory is headed downhill, 
an entire sequence of $E$, corresponding to transient configurations,
 may arise which is unrelated to landscape minima.
Only the $E$ value preceding a 
barrier record has potential physical significance, since
it will again    be
repeatedly visited and is likely associated to 
a new local equilibrium situation. 
Similarly, the sequence of barrier records 
of roughly the same magnitude which may appear 
while a trajectory crosses high `saddles' is 
uninteresting, except for the 
highest barrier record preceding a
new low energy minimum.

Consider now a sequence of energy and barrier 
records, e.g.\ the $EBBEEBBBE$ shown 
in Fig.~\ref{fig1}. As argued, we only keep
the full squares and circles (the `$\otimes$' symbol
represents a datum removed) thereby reducing our string 
to its final state: $E(BEB)E$. 
 Within the triplet $(BEB)$ the two $B$'s identify,
 by definition, the entry and 
exit times of a valley, and the $E$ identifies the time $t_{hit}$ at 
which the lowest state of the valley is hit for the first time. The 
recurrent quasi-equilibrium fluctuation regime starts at $t_{hit}$ 
and lasts until the valley is exited. 
Energy minima which are not records 
 are not assigned a separate valley, and will
create an internal structures within each valley. Finally,
no new valleys can ever appear after the ground state
is discovered. 
 
Our  protocol is maximally restrictive 
in its definition of a new valley: different landscape minima 
are lumped together whenever either they  or the    
 energy barriers separating   them are degenerate. 
Even   models whose aging purely
originates from entropic sources~\cite{Leuzzi02} 
would only produce one valley.
In the spin 
glass energy landscape the approach nonetheless uncovers a great 
deal of structure,  whose dynamical significance
is more fully discussed in Refs.~\cite{Sibani03,Dall03}.

For simulation speed, we rely on the Waiting Time Method
(WTM), a rejectionless Monte Carlo scheme~\cite{Dall01} 
well suited for problems where $N$ variables
contribute additively to the energy through
 local interactions. Based on the local field,
each variable is stochastically assigned a 
flipping time and the variable with the shortest flipping time is 
updated together with the local fields affected by the move.
The sequence of WTM moves equals in probability~\cite{Dall01} 
 the sequence of accepted moves in the Metropolis algorithm. 
The current flipping time, henceforth
simply `time', corresponds to Metropolis sweeps,
as well as to the physical time of an experiment.

\section{Model} We consider $N$ Ising spins 
interacting via the Edwards-Anderson 
Hamiltonian~\cite{Edwards75}
\begin{equation}
{\cal H}(\alpha) = -
\frac{1}{2}\sum_{i,j} J_{ij}s_i^{(\alpha)}s_j^{(\alpha)}.
\end{equation} 
In this formula, $s_i^{(\alpha)}$
 is the spin value at site $i$ 
for configuration $\alpha$, 
the couplings $J_{ij}$ are
symmetric in their two indices and vanish 
unless $i$ and $j$ belong 
to neighboring sites. In the
latter case they are 
drawn independently 
from a Gaussian distribution of unit variance.
We mainly consider regular lattices of linear extension $L$ in 
 three dimensions, where $N=L^3$.
The dynamics is single spin flip with 
 detailed balance. 

\section{Results} 
\begin{figure}[t]
\twofigures[scale=0.37]{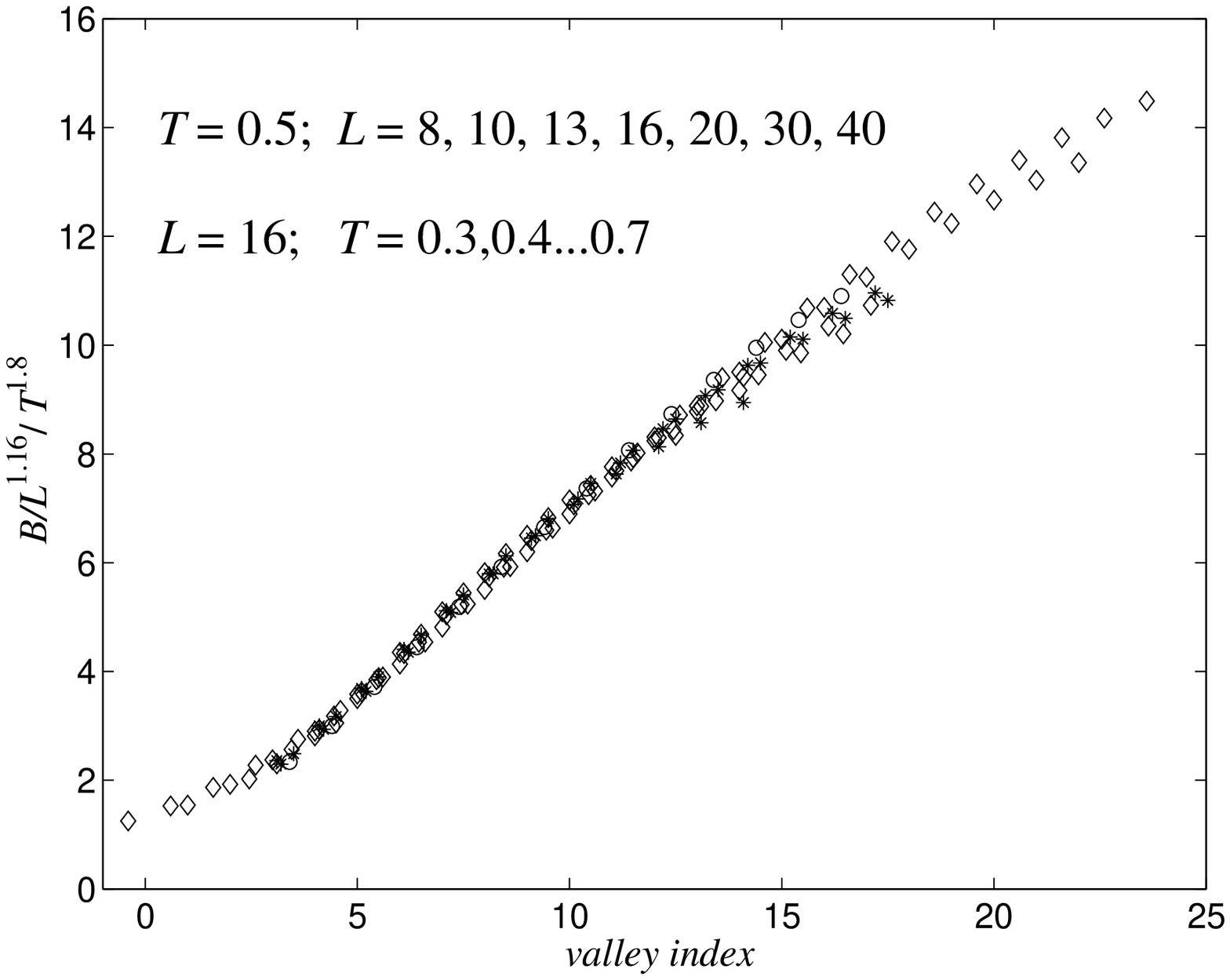}{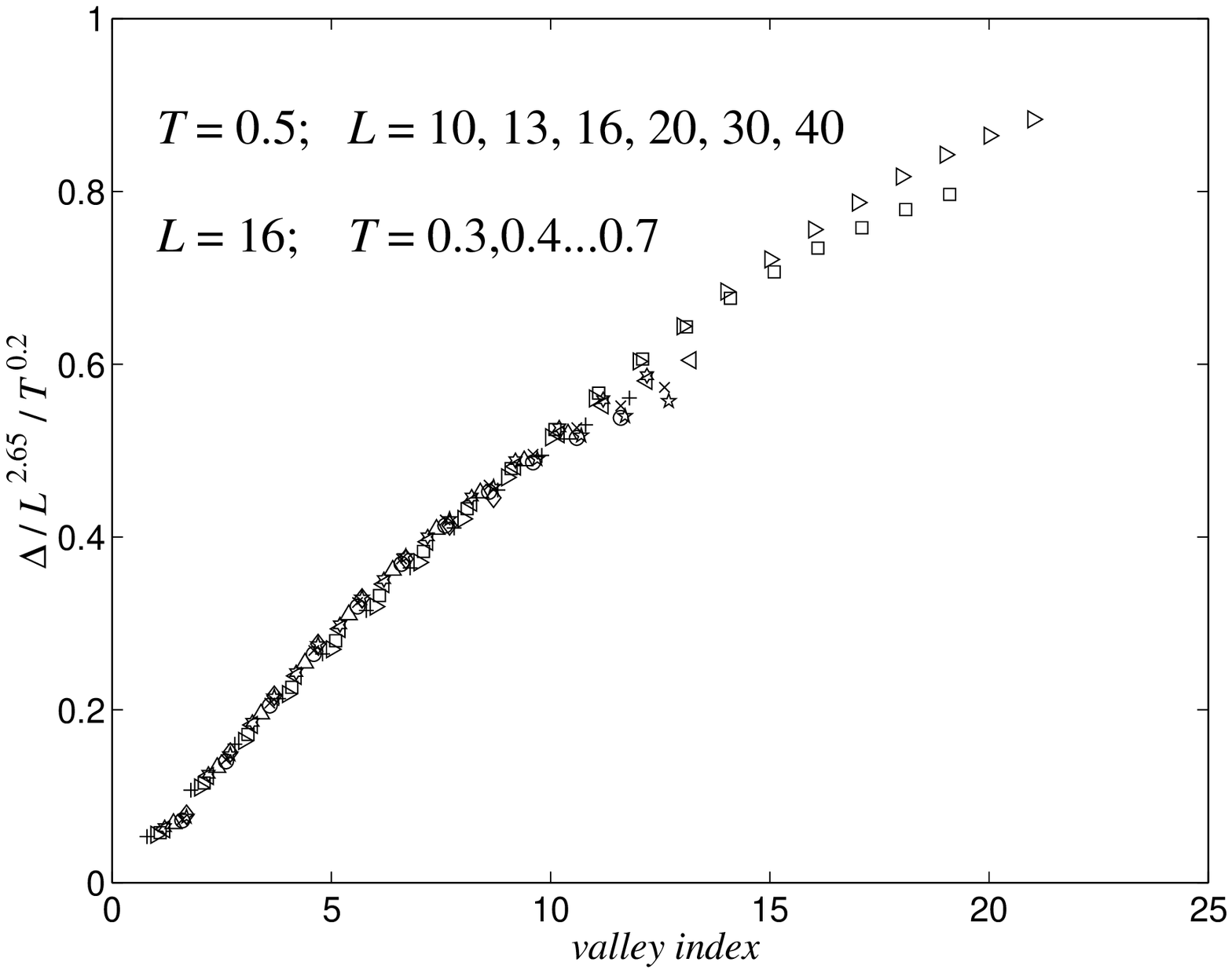} 
\vspace{-0.2cm}
\caption{\small The average size of the barrier $B$ separating contiguous 
valleys in $3d$ spin glass landscapes is scaled with $L$ and $T$ as indicated 
in the ordinate label and plotted versus the valley index $n$.
The combinations of $L$ and $T$ values specified in the figure were included
in the scaling plot. The abscissa pertaining to different data sets is shifted 
by amounts of order one.}
 \label{scaling_B}
 \caption{\small The average
difference $\Delta$ between the lowest energies of the first valley 
and the $n$'th valley 
 is plotted versus the valley index $n$. The ordinate is scaled with $L$ 
as indicated, and the abscissa pertaining to different data sets is shifted 
by amounts of order one.  The small $T$ dependence of  $\Delta$ implies that
nearly  the same energy is lost through $n$ quakes  at different temperatures,
even though the barriers overcome widely differ.}
\label{fig4}
\end{figure}

To obtain meaningful 
 comparisons, all data originating from the 
 first ten time units after the quench were discarded.
 Secondly, the   index identifying
 each valley within the ordered sequence of
 valleys visited---for short valley index---was shifted
  up to two units in Figs.~\ref{scaling_B}~and~\ref{fig4}.
 Unless otherwise specified,  our data are quenched averages 
 over an ensemble, whose size 
 varies from $30 000$ for the smallest systems 
 to $4000$ for the largest ones. In all cases,
the amount of statistics was sufficient to remove any visible scatter. 
The systematic error left stems from the finite 
run time coupled to the broadness of the distribution of the 
intervals between successive quakes. This leads 
to a considerable variation 
in the number of valleys seen in different runs. 
Data corresponding to valley indices achieved by less than 
$75$\% of the ensemble were discarded to reduce the error.

A basic quantity is the average number $\overline{n_V}(t)$ of quakes
 or valleys plotted in 
Fig.~\ref{B_in_time} as a function
 of $\ln t$. Note that the smooth appearance 
 of the curves stems from the large amount of data
 available, rather than from fitting. 
 All the plots in the main figure belong to 
 $3d$ systems of growing size, while similar data in $2d$ 
 and for smaller systems were omitted for clarity. 
 The insert shows the logarithmic rate of events $\alpha = d \overline{n_V}(t)/ d \ln t$, 
in $2d$ as well as $3d$, as a function of 
$N=L^d$. The derivative was calculated, ignoring 
the small systematic curvature, as the average slope between 
times $10^4$ and $10^6$. Notably,  $\alpha$ is highly insensitive 
to the  temperature, and to the system dimension. 
The variance of $n_V$ was also calculated and its ratio to the
average found to be constant, except at short times. Thes properties
are accounted for by 
a \emph{log-Poisson}
statistical description~\cite{Sibani93a}, whose connection
to glassy dynamics is developed in Ref.~\cite{Sibani03}.
 
The statistics implies that the energy function decorrelates 
 between consecutive quakes. Since 
 the correlation time and relaxation time 
(for the internal thermalization) can be 
generically identified~\cite{VanKampen92}, 
quasi-equilibrium is typically achieved before the valley is exited.

 Conversely, whenever dynamical trajectories  dwell
 near a local minimum,
record high energy fluctuations decorrelate, 
and, hence, a log-Poisson distribution applies to them~\cite{Sibani93a}.
Crucially, the definition of a quake  additionally requires 
 the achievement of a record \emph{low} energy value. The availability
 of such minima  mirrors
 the landscape geometry and the selection of the 
 initial attractor. E.g.,  
 if the trajectories were started at the ground state, the system
 would for ever remain in the same valley.
The fact that the same statistical description covers
 barrier records as well as quakes shows that 
 climbing the former triggers the latter. 
The logarithmic rate of events grows with the 
lattice size $N$ as shown in the 
insert of Fig.~\ref{B_in_time}, i.e.\ in 
a nearly logarithmic
fashion, from an initial value close to $1$ to a value 
close to $2$ for extremely large system sizes. 
For a Poisson process 
this rate is proportional to
the number of lattice `regions' evolving independently~\cite{Sibani03},
and its reciprocal is therefore proportional to the size
in real space of the region affected by a quake.
Importantly, the logarithmic nature 
of the statistics implies that the time spent to reach the bottom of a valley 
entered at a time $t_w$ is
 proportional to $t_w$~\cite{Sibani03}, whence
 the near-extensive  configurational and energy changes
we observe  are not the 
 outcome of instantaneous processes.

Further insight into the properties of quakes
 can be gained through the scalings shown in Figs.~\ref{scaling_B} and~\ref{fig4}.
Figure~\ref{scaling_B} describes the scaling with $L$ and $T$ 
 of the average energy barrier $B$ separating two
contiguous valleys, while Fig.~\ref{fig4} 
 depicts the average difference $\Delta(1,n) = E_1 - E_n$ between the
state of lowest energy in valleys $1$ and $n$. 
 While we have not attempted
a quantitative determination of the uncertainty 
on the scaling exponents, changing the last significant 
digit visibly affects the quality of the collapse. 
 
The smooth and nearly linear increase of 
the energy barrier with $n$ confirms 
the expectation that the valleys gradually become more
 stable against thermal fluctuations. 
The growth of the barriers with system size is $\propto L^{1.16}$, while
 the typical range of the energy fluctuations is
 $\propto L^{3/2}$. While the slower growth observed is consistent with 
 a bias favoring low energy saddles, 
 the strong temperature dependence $\propto T^{1.8}$
 clearly indicates an entropic effect: A linear 
 $T$ scaling would apply if the saddle of lowest energy
were the only route  utilized. The average Hamming distance $H_d$ between the bottom states
of contiguous valleys (see Ref.\cite{Dall03} for a figure) 
was also investigated
and found to scale with $L^{2.85}$ and $T^{1.70}$.

Turning to Fig.~\ref{fig4}, the initially linear increase 
of $\Delta(1,n)$ is seen to slowly taper off, possibly reflecting 
a change in landscape geometry closer to the ground state.
The $L^{2.65}$ dependence of $\Delta(1,n)$
 also characterizes  its derivative with respect to $n$, i.e.\ 
the energy difference $\Delta(n, n+1)$ between neighbor valleys.
 
The weak temperature dependence of the scaling plot
 is to be contrasted with the strong $T$
 dependence of both the barriers overcome and of the
 Hamming distance. After $n$ quakes, two systems relaxing 
 at different temperatures lose nearly the same 
 amount of energy, even though the configurations visited 
 and the energy barriers crossed substantially differ. 
 Firstly, this implies a strong degeneracy of the minima.
 Secondly, in conjunction with the striking 
 insensitivity to the temperature of the time statistics 
 of the quakes~\cite{Sibani03,Dall03}, it implies that 
 the dominant paths  at different $T$ widely differ.\ 
 Qualitatively, this idea appeared early 
 in the spin-glass literature~\cite{Dotsenko85}, and was
 later incorporated in a model~\cite{Hoffmann97}
 showing rejuvenation effects~\cite{Jonason98}.
In a broad  sense, our findings concur with the view of aging
as activated dynamics within a hierarchy 
of energy barriers~\cite{Sibani89,Lederman91,Sibani91,Joh96,Hoffmann97,Sibani97a,Joh99}. 
 \begin{figure}[ht!]
\twofigures[scale=0.37]{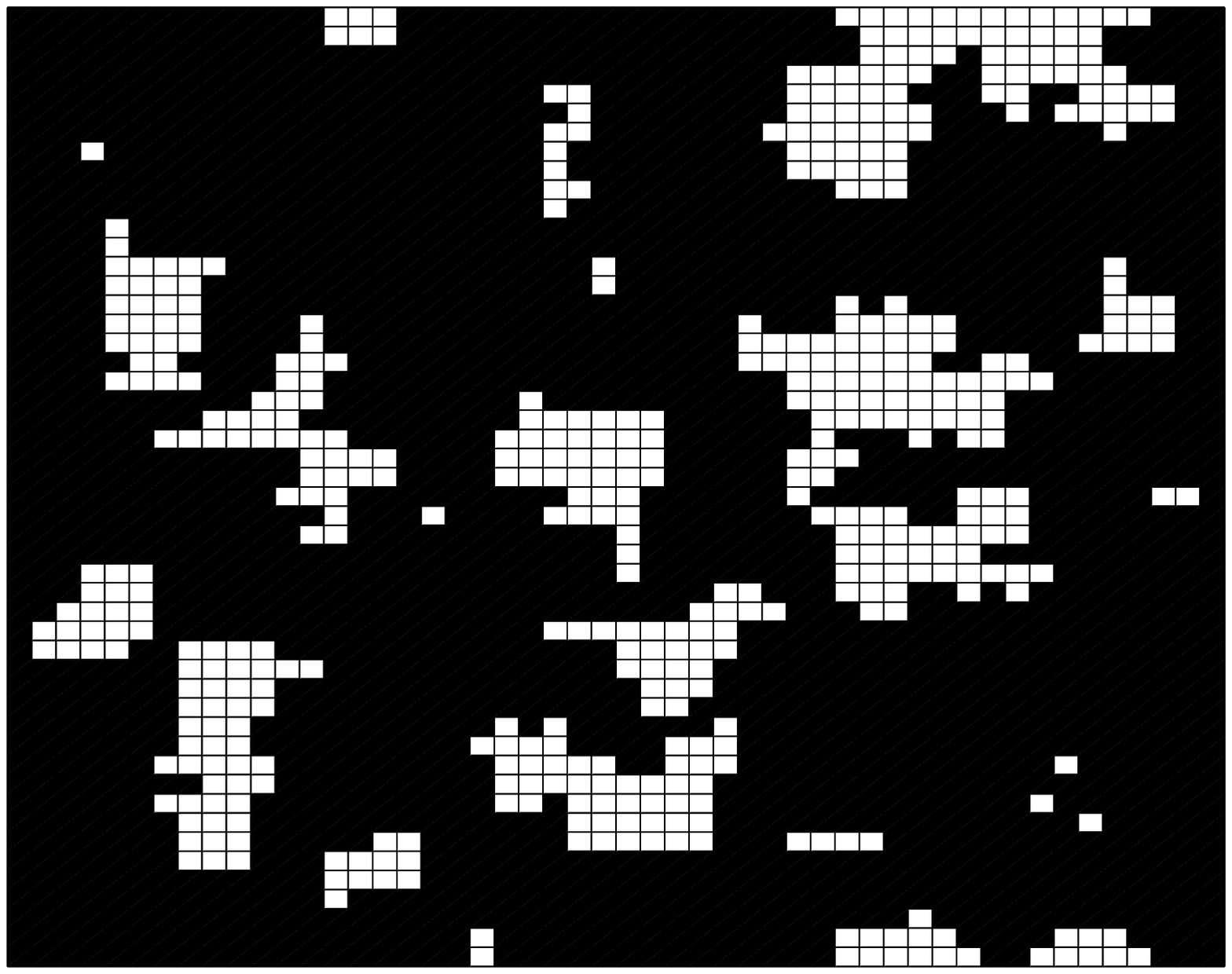}{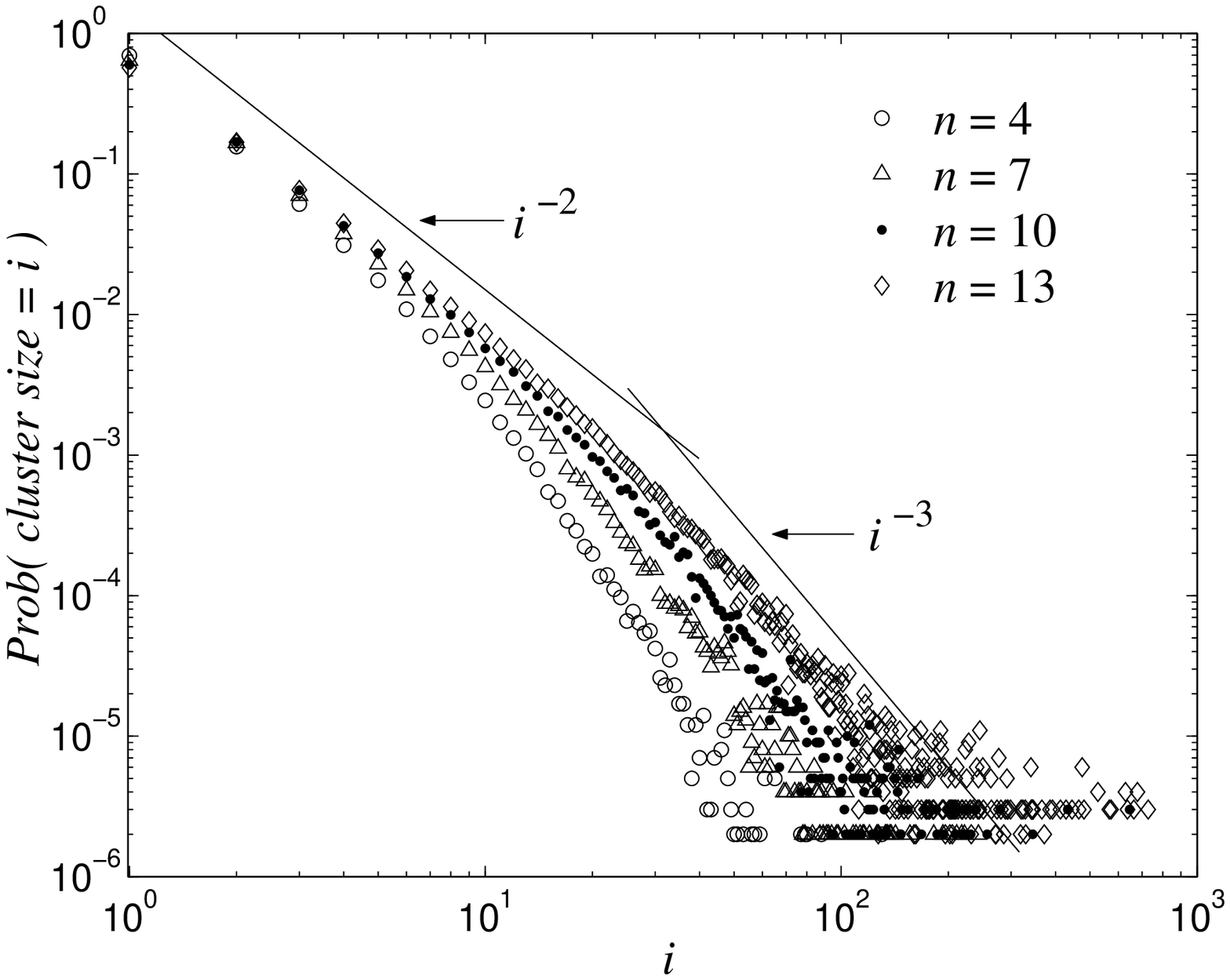} 
\caption
{ \small The    spins overturned by the $n$'th
quake  are shown in white  
in the  projection of the two lowest-lying states of 
valleys $n$ and $n+1$ onto each other. The data pertain to 
a $2d$ system of size $N=50^2$ at $T=0.3$, with $n=21$. 
}
\label{fig5} 
\caption{ \small The distribution of cluster sizes for a $3d$
system of linear size $L=30$ is shown for a sequence of valley 
indices $n$.\ The simulation temperature is $T=0.4$. The straight
lines corresponding to power laws are included as a guide to the eye. 
}
\label{fig6} 
\end{figure} 

As shown in Fig.~\ref{fig5}, spins flipped through a quake 
fall into $m$ connected clusters, whose surface 
 carries the corresponding 
non-equilibrium energy release\footnote{Quite unlike 
 droplets~\cite{Fisher88a,Koper88}, which carry the energy
\emph{accumulated} through a quasi-equilibrium fluctuation.}. 
Letting $m(i)$ denote the number of clusters of size $i$ within
a given system and 
$p(i) = m(i)/m$, we average $p(i)$ over $1000$ 
realizations of the couplings $J_{ij}$ to obtain the 
distribution of cluster sizes shown in Fig.~\ref{fig6}
for a series of quake indices $n$. 
As a guide to the eye, straight lines are included,
representing power-laws with exponents 
$-2$ and $-3$. 
As the system ages, the cluster size distribution broadens, with the early decay 
close to the $i^{-2}$, and with the  the tail of
the distributions falling off more slowly   than  
$i^{-3}$, except for small $n$.
   This means that  the variance  of the cluster size  is dominated by the end of the
integration interval and thus grows with $N$. Regardless of whether the
average cluster size might remain  finite or not in the limit of 
large $n$ and large $N$, the average cluster is never
the typical one. As  Fig.~\ref{fig6} more
than suggests,   the latter    diverges with the system size.  
Also note that, were this not the case,   
the size  dependence of the energy and configuration changes
 $\Delta(n, n+1)$ and $H_d(n)$ would both scale with the 
 number of clusters present in the system.  $\Delta $ and $H_d $ would  
 then possess identical $N$ scalings, which  
 contradicts  our previous findings.
   
A direct study of the size dependence of the distribution
including even larger sizes 
is exceedingly difficult to carry out. In order 
to avoid comparing a larger  but  younger  system with 
one which is   smaller but older, the quake index $n$
must be increased    logarithmically as $N$ 
increases, see e.g.\ Fig.~\ref{B_in_time}.

\section{Conclusions and outlook} 
The gist of our approach is to uncover 
the dominant structural 
landscape features, e.g.\ barriers, minima and 
distances, as they appear to 
the unperturbed aging dynamics
far from the ground state and far from global 
thermal equilibrium. We strongly emphasize 
the qualitative difference between 
local equilibrium fluctuation dynamics within 
a valley and the non-equilibrium  quake dynamics
through which the   energy  trapped 
by the initial  quench is slowly released. 
 The latter process,
which is practically irreversible
on the time scales considered, has  not
received much theoretical attention. 
A recent effort~\cite{Sibani03}
describes it   
in terms of a log-Poisson process, 
establishing a  connection  to 
 driven dissipative dynamics~\cite{Sibani93a,Sibani01} and
evolution models~\cite{Sibani99a,Hall02}, where 
 the pace of the dynamics similarly 
decelerates while the system ages.
 
Uniquely to our landscape approach,
 the resolution of the probe 
self-adjusts to detect gradually coarser scales
as the pace of the dynamics slows down.
The picture with emerges 
confirms the presence of a hierarchy 
of dynamical barriers which must be overcome 
in order for the system to relax, and   
 complements the results 
obtained by other methods.
In the diffusive dynamics of atomistic
glassy models above the glass transition temperature
studied by Doliwa and Heuer~\cite{Doliwa03}, 
 energy barriers also  play a decisive 
role, and `metastructures', which  similarly  to our valleys
comprise many local minima labeled by their
lowest energy state,  are the main
objects in a coarse grained dynamical description. 
In  short range spin-glasses, the  morphology of local energy 
minima  with energy of order one
relative to the ground
state   was investigated by Krzakala,
 Martin and Houdayer~\cite{Krzakala00,Houdayer00}. 
 These states  differ  from
the ground state on a set of spins spanning the 
 whole system,   which clearly 
 separates them  from droplet-like 
 excitations and suggests a  similarity to the 
 quake-induced rearrangements shown in  Fig.~\ref{fig5}. 
 Clearly, the dynamical regime presently considered 
 remains quite far from the
 ground state, and the energy difference are also far 
 being of order one.  In the extreme asymptotic regime
 where quakes become increasingly rare 
 even on a \emph{logarithmic} time scale---a trend 
 visible for small systems in 
 Fig.~\ref{fig1}---the energy released 
 must however approach zero. If the trend expressed by Fig.~\ref{fig6}
 continues in this regime, our dynamically
 identified clusters might then have morphological properties similar to 
 the `sponges' in Refs.~\cite{Krzakala00,Houdayer00}.
 
Let us finally consider the
connection to   \emph{pseudo-equilibrium} thermal correlations~\cite{Berthier02}.
In short-ranged spin-glasses these   are established   
 on an age dependent  length scale $l(t_w)$ which  can be  extracted
 from a four point thermal correlation function. This scale  \emph{i)}  grows in a slow power-law fashion  
 with  the age of the system, and \emph{ii)} stays below $l(t_w)\approx 4$ in the range of temperature
 and times presently considered. Since   four point correlations are insensitive
 to a coherent  spin flip  occurring on a scale much larger than $l(t_w)$,
 and since   equilibrated clusters or droplets easily 
 fit inside the much larger quake-induced rearrangements implied  by Fig.\ref{fig6},
  the non-equilibrium de-excitation mechanism presently  described seems only  
 weakly, if at all, coupled to the local thermalization process. This is
 completely   in line with the approach of Ref.~\cite{Sibani03}.

\vspace{0.2cm} 
\noindent {\bf Acknowledgments}:
The authors are indebted to Greg Kenning and Henrik J. Jensen for discussions, 
to the Danish SNF for grant 23026, and to the Danish Center for Scientific
Supercomputing (DCSC) for computer time on the Horseshoe Linux cluster in Odense.



\end{document}